\documentstyle[aps,prl,multicol,floats,epsf]{revtex}
\epsfclipon
\newcommand{\smeq}{\! = \!}

\newcommand{\lefthalfline}
   { \vspace*{-0.2truein} \noindent \hrulefill \hspace*{3.6truein} }
\newcommand{\righthalfline}
   { \vspace*{-0.0truein} \hspace*{3.6truein} \hrulefill \vspace*{0.0truein} }

\draft

\newcommand{\bc}{\begin{center}}
\newcommand{\ec}{\end{center}}

\newcommand{\be}{\begin{equation}}
\newcommand{\ee}{\end{equation}}
\newcommand{\ba}{\begin{array}}
\newcommand{\ea}{\end{array}}

\newcommand{\cl}{|\chi_{\rm Landau}|}
\newcommand{\kf}{k_{\scriptscriptstyle F}}

\newcommand{\Ef}{E_{\scriptscriptstyle F}}

\newcommand{\br}{{\bf r}}

\newcommand{\G}{{\cal G}}   

\begin{document}



\title{\bf Interaction-Induced Magnetization of the Two-Dimensional 
Electron Gas}

\author{ Felix~von~Oppen,$^{(1)}$ Denis~Ullmo,$^{(2)}$ and
Harold~U.~Baranger$^{(3)}$ }

\address{$^{(1)}$Institut f\"ur Theoretische Physik, Z\"ulpicher Str.\ 77,
Universit\"at zu K\"oln, 50937 K\"oln, Germany}

\address{$^{(2)}$Laboratoire~de~Physique~Th\'eorique~et~Mod\`eles~%
Statistiques~(LPTMS),~91405~Orsay~Cedex,~France}

\address{$^{(3)}$Department of Physics, Duke University, Box 90305,
Durham, NC 27708-0305}

\date{Submitted to Phys. Rev. B, 7 January 2000}
\maketitle

\begin{abstract}
We consider the contribution of electron-electron interactions to the
orbital magnetization of a two-dimensional electron gas, focusing on the
ballistic limit in the regime of negligible Landau-level spacing. This
regime can be described by combining diagrammatic perturbation theory with
semiclassical techniques. At sufficiently low temperatures, the
interaction-induced magnetization overwhelms the Landau and Pauli
contributions. Curiously, the interaction-induced magnetization is
third-order in the (renormalized) Coulomb interaction. We give a
simple interpretation of this effect in terms of classical paths using a
renormalization argument: a polygon must have at least three sides in
order to enclose area. To leading order in the renormalized interaction,
the renormalization argument gives exactly the same result as the full
treatment.
\end{abstract}







\widetext
\begin{multicols}{2}

\section{Introduction}

Within the independent-electron picture, the magnetic response of a bulk
two-dimensional electron gas has two sources: Pauli paramagnetism originating
from the electron spin and Landau diamagnetism originating from the orbital
electronic motion. After studies of the contribution of superconducting
fluctuations to the magnetic response of superconductors above $T_c$
\cite{lee,eilenberger}, Aslamazov and Larkin \cite{aslamazov} pointed out
that electron-electron interactions make an analogous contribution to the
magnetic response of normal-metal systems. While the fluctuation contribution
is diamagnetic in superconductors, the Coulomb interaction gives a
paramagnetic contribution to the susceptibility of normal metals; this
difference is a direct consequence of the different signs of the effective
interaction in the two cases.

In their seminal work, Aslamazov and Larkin \cite{aslamazov} computed the
interaction contribution to the susceptibility of three-dimensional metals
and of layered systems at zero magnetic field. They found that the effect was
particularly strong for layered systems. In view of the importance of the
physics of the two-dimensional electron gas, the purpose of the present paper
is to compute the interaction-induced magnetization of a strictly {\em
two-dimensional} bulk system. We shall moreover go beyond the zero-field
limit considered by Aslamazov and Larkin and compute the magnetization for
arbitrary classically weak magnetic fields. We find that the
interaction-induced magnetization generally dominates over the Landau and
Pauli contributions at sufficiently low temperatures.

The relevant length scales of the problem are the thermal length $L_T \smeq \hbar
v_F / (2\pi T) $, the magnetic length $L_H \smeq (\hbar/eB)^{1/2}$, the cyclotron
radius $R_c \smeq m_ev_F/eB$, and the elastic mean free path $\ell_{el}$.
Throughout this paper, we focus on the regime $L_T \ll \ell_{el}$, which
allows us to neglect the effects of impurity scattering. Moreover, we
restrict ourselves to classically weak magnetic fields, defined by the
condition $L_T\ll R_c$ (or equivalently $\hbar\omega_c\ll T$, where
$\omega_c$ is the cyclotron frequency). Within a semiclassical approach,
this implies that we can neglect the classical effects of the magnetic field
on the trajectories and only need to consider the Aharonov-Bohm phases
induced by the $B$ field. For classically weak magnetic fields, we can
distinguish between two magnetic field regimes: The low-field regime,
considered by Aslamazov and Larkin, where $L_T\ll L_H$ and the high-field
regime $L_T\gg L_H$. We present analytical results in both regimes and also
show numerical results bridging these two regions. Most importantly, we
present a simple renormalization argument based on classical paths which
exactly reproduces the result of the complicated full treatment.

The magnetization of the two-dimensional electron gas has been studied
experimentally in mesoscopic samples \cite{Levy2,Benoit} and in the
quantum-Hall regime \cite{Eisenstein,Usher,Levy3,Heitmann}. To the best of
our knowledge, no experiments have been performed on bulk 2d samples at
classically weak magnetic fields. Such experiments would be a valuable test
of our theoretical understanding of the interaction contribution to the
magnetization.

This paper is organized as follows. In Sec.~\ref{sec:semi} we develop the
semiclassical approach to the interaction-induced magnetization for 2d bulk
systems. In Sec.~\ref{sec:cal} we employ the general results derived in
Sec.~\ref{sec:semi} to derive explicit expressions for the magnetization in
the high- and low-field regimes. A curious feature of these results is that
the interaction-induced susceptibility is third order in the (renormalized)
interaction strength. In Sec.~\ref{sec:RG}, we show how the renormalization
group approach introduced in \cite{prl} allows one to give a simple
semiclassical interpretation of this result. We conclude in Sec.\
\ref{sec:con} by comparing the interaction-induced susceptibility to the
Landau and Pauli susceptibilities and discussing finite-size effects.

\section{The semiclassical approach}
\label{sec:semi}

\subsection{Basics}

\subsubsection*{Cooper-channel}

Calculating the interaction contribution to the magnetic response
requires one to extend the high-density expansion (RPA) of the
thermodynamic potential \cite{AGD} by including interaction
corrections from diagrams in the Cooper-channel.  This was first done
in the context of superconducting fluctuations and then applied to
disordered normal
metals\cite{aslamazov,AltArZu,doubleA,ambegaokar,Eckern}.  Such
expansions usually give reliable results even beyond the high density
limit, if the relevant sets of terms are properly resummed.  The
relevant Cooper-like diagrams are shown in Fig.\ 1.  The screened
Coulomb interaction (wavy lines) can be treated as
local\cite{doubleA,ambegaokar}: $U(\br - \br')  \smeq \lambda_0 N(0)^{-1}
\delta (\br - \br')$. Here, $N(0) \smeq m_e/(\pi\hbar^2)$ denotes the full
density of states and the bookkeeping index $\lambda_0\!=\!1$
identifies the order of perturbation. For a local interaction,
the direct and exchange term are the same up to a factor of $(-2)$ coming
from the spin sums and the different number of fermion loops.  
The  straight lines in Fig.~1
represent finite-temperature  Green functions of the non-interacting system.
These take the form
\begin{equation} \label{Gtemp}
\G_{\br, \br'}(\epsilon) = \theta (\epsilon)G^R_{{\bf r},{\bf r'}}
(\Ef \!+\!i\epsilon) + \theta(-\epsilon) G^A_{{\bf r}, {\bf r'}}
(\Ef \!+\!i\epsilon)
\end{equation}
in terms of the retarded and advanced Green functions
$G^R_{\br,\br'}(E)=[G^A_{\br',\br}(E^*)]^*$.

The perturbation expansion for this interaction contribution $\Omega$ to
the thermodynamic potential, which yields the magnetic response, can
be formally expressed as \cite{AltArZu,doubleA}
\end{multicols}
\lefthalfline
\begin{eqnarray}
      \Omega & = &  -{1\over \beta} \sum_{n=1}^\infty
       \frac{(-\lambda_0)^n}{n} \,
       \sum_\omega \int d{\bf r}_1 \ldots d {\bf r}_n \,
       \Sigma_{\omega}(\br_1, \br_2) \ldots
       \Sigma_{\omega}(\br_n, \br_1)   
                                \label{omega_expanded}\\
       & = & {1\over \beta} \sum_\omega {\rm Tr} \left\{
       \ln [1+ \lambda_0 \hat \Sigma_\omega ] \right\} \;. 
                                \label{omega_diag} 
\end{eqnarray}
\righthalfline
\begin{multicols}{2}\noindent
Here, $\omega$ denotes the bosonic Matsubara frequencies $\omega \smeq 2\pi
m/\beta$ ($m$ is any positive or negative integer) with $\beta \smeq  
1/k_B T$.  (We employ units such that $k_B \smeq 1$ in this paper.)  The
particle-particle propagator $\Sigma_\omega$ is expressed (in
position representation) in terms of products of finite--temperature
Green functions as\cite{AGD}
\begin{equation} 
\label{sigma}
        \Sigma_{\omega}(\br, \br')={1\over \beta N(0)}
        \sum_{\epsilon}^{\Ef} {\cal G}_{{\bf r}, {\bf r'}}
        (\epsilon) {\cal G}_{{\bf r}, {\bf r'}}(\omega-\epsilon)
\end{equation}
where      the sum    runs    over  the     fermionic Matsubara    frequencies
$\epsilon=(2n+1)\pi / \beta$.  The short-length  (high-frequency) behavior  is
included in the screened interaction, thus requiring a cutoff of the frequency
sums at the Fermi  energy ${\Ef}$\cite{doubleA}. 

\subsubsection*{Semiclassical Green function}

In view of the fact that the Fermi wavelength is the smallest length
scale in the problem, our strategy will be to replace the free Green
functions in the particle-particle propagator by their semiclassical
approximations.  Generally, the semiclassical approximation to the 
Green function
$G_{{\bf r},{\bf r'}} (\Ef \!+\!i\epsilon,B)$ is expressed as a sum over
all classical paths from ${\bf r}$ to ${\bf r'}$ at energy $E_F$
\cite{Gutzwiller}.  For the bulk 2d electron gas, there is only a
single such trajectory, namely the straight line connecting the two
points.  For $\epsilon=0$ and $B=0$  one therefore has 
\begin{equation}
  G^R_{{\bf r},{\bf r'}}(\Ef,B\!=\!0) \simeq 
          \frac{1}{ i \sqrt{2\pi}} \frac {m_e}{ \hbar^2}
  \frac {\exp[iS_{{\bf r},{\bf   r'}}(E_F)/\hbar-i\pi/4] }
              {(k_F|{\bf r}-{\bf r'}|)^{1/2}}
\end{equation}
where $S_{{\bf  r},{\bf r'}}(E_F) = \hbar  \kf  |{\bf  r}-{\bf r'}|$   is the
classical action along the path.  Moreover, since we assume the magnetic
field is classically  weak, the field affects the action along the path
through
\be 
   S(B) =  S({B\!=\!0}) +  {e \over c}
   \int_{\bf r}^{\bf r'} d{\bf x}\,{\bf A} ({\bf x})
\ee
where the integral is along the unperturbed straight line path.
Finally, it turns out  that   only small  values    of the
imaginary part  of  the  energy  $\epsilon$ should be considered, so
we can use the relation
\be
   (\partial S / \partial E) = t \; ,
\ee
where $t = |{\bf r}-{\bf r'}| / v_F$ is the time of flight from ${\bf r}$ 
to ${\bf r'}$.
In this way, one obtains the semiclassical Green function for finite field
and finite (imaginary) energy $\epsilon$:
\end{multicols}
\lefthalfline
\begin{equation}
  G^R_{{\bf r},{\bf r'}}(\Ef \!+\!i\epsilon,B)=
  G^R_{{\bf r},{\bf r'}}({\Ef,B\!=\!0})
  \exp\left[ {ie \over \hbar c} \int_{\bf r}^{\bf r'} d{\bf x}\,{\bf A} 
  ({\bf x}) \right]
  \exp\left[-{|\epsilon||{\bf r}-{\bf r'}| \over \hbar v_F } \right] \; .  
\end{equation}
\righthalfline
\begin{multicols}{2}

\subsubsection*{Semiclassical particle-particle propagator}

In the calculation of the thermodynamic potential Eq. (\ref{omega_expanded})
one should neglect all rapidly oscillating contributions in 
$\Sigma_{\omega}(\br, \br')$
as these will give a small contribution upon integration. Thus, in the
particle-particle propagator $\Sigma$, it is necessary to pair
advanced and retarded Green functions, and furthermore to pair each path in
the semiclassical expression for $G^R$ with those in $G^A$ for which the
dynamical phase factor cancels.  The obvious case of pairing each path with
itself is excluded because it yields no field dependence in $\Omega$ and
hence zero magnetization.  Thus one is led to consider pairs of
time-reversed paths-- for these the dynamical phase cancels but the magnetic
field part is multiplied by two.  The pairing of $G^R$ with $G^A$ means
concretely that one should keep only those terms  in which $\epsilon$ and
$\omega - \epsilon$ have opposite sign in the sum over Matsubara frequencies.
Using the relation
\be
  \sum_{\epsilon (\omega - \epsilon) < 0} \exp [- |2\epsilon-\omega| t /\hbar]
  =   \frac{\exp[- |\omega| t / \hbar]}{ \sinh (2 t \pi / \beta \hbar)} \; ,
\ee
one obtains the final result for $\Sigma$
\be \label{eq:sigma_sc}
  \Sigma_\omega({\bf r},{\bf r'}) =
  \Sigma_\omega^{(0)}(|{\bf r}-{\bf r'}|)  
  \exp\left\{{2ie\over c\hbar}\int_{\bf r}^{{\bf r}^\prime}
       d{\bf x}\,{\bf A} ({\bf x})\right\} \nonumber 
\ee
where we have introduced the zero field limit of the particle-particle
propagator
\be \label{eq:sigma0}
  \Sigma_\omega^{(0)}(|{\bf r}-{\bf r'}|) = 
                \frac{1}{4\pi L_T |{\bf r}-{\bf r'}| }
    \frac{\exp[- |m| |{\bf r}-{\bf r'}| / L_T]}
    { \sinh (|{\bf r}-{\bf r'}|  / L_T)} \; .
\ee

\subsection{Derivation of the eigenvalues}

The ladder-diagram  contribution  to the thermodynamic  potential is
expressed in Eq.~(\ref{omega_diag}) solely in terms of the eigenvalues
of the operator $\Sigma_\omega({\bf r},{\bf r}^\prime)$.  We therefore need
to solve the eigenvalue equation
   \begin{equation} \label{eq:eigen_eq}
        \int d{\bf r^\prime} \, \Sigma_\omega({\bf r},{\bf r}^\prime)
     \psi_{n,l}({\bf r}^\prime)=\sigma_\omega^{n,l}\psi_{n,l}({\bf r})
   \end{equation}
where $n$ and $l$ are quantum numbers.

Assuming from now on the symmetric gauge  ${\bf A}={\bf B}\times{\bf r}/2$,
Eq.~(\ref{eq:sigma_sc}) reads
\be   \label{eq:Bdep}
  \Sigma_\omega({\bf r},{\bf r}^\prime) =
  \Sigma_\omega^{(0)}(|{\bf r}-{\bf r}^\prime|)
  \exp\left\{{2ie\over c \hbar} {B\over 2} ({\bf r} \times {\bf r'}) \right\}
  \; .  
\ee
It can be easily checked that any operator of the form Eq.~(\ref{eq:Bdep})
commutes with any element of the magnetic translation group
  \begin{equation}
  \hat T({\bf R}) = \exp \left({i \over \hbar} {\bf R} \cdot 
               (\hat {\bf p} - {2 e \over c} {\bf A}) \right)
   \end{equation}
and therefore with its generators
  \begin{eqnarray*}
  \hat \Pi_x & = & (\hat p_x -(2e/c)  A_x) \\ 
  \hat \Pi_y & = & (\hat p_y -(2e/c)  A_y) \; .
  \end{eqnarray*}
Noting that, first, $\Sigma_\omega({\bf r},{\bf r}^\prime)$ is invariant
under rotation and, second, the Landau Hamiltonian for a particle of charge 
$(-2e)$ can be written as
   \begin{equation}
   \hat H_L = \left (\hat {\bf p} + {2 e \over c} {\bf A} \right)^2
       = \hat \Pi_x^2 + \hat \Pi_y^2 + {4 e \over c \hbar^2} \hat J_z \; ,
   \end{equation}
we see that
$\Sigma_\omega({\bf r},{\bf r}^\prime)$ is diagonal in the basis
$\left\{ \psi_{n,l} \right\}$ of the eigenvectors of $\hat H_L$ and $\hat J_z$, 
where $n$ and $l$ are the Landau level and angular momentum quantum 
numbers, respectively. For $l \smeq 0$ the Landau level wavefunction
(for a particle of charge $-2e$) has the well-known form
\begin{equation} 
  \psi_{n,0}({\bf r}) = \exp\left\{- {|{\bf r}|^2 \over 2
  L_H^2} \right\} L_n\left( {|{\bf r}|^2 \over L_H^2} \right)
\end{equation} 
with $L_n$ the Laguerre polynomial and $L_H = (\hbar/eB)^{1/2}$.
   
Finally, an important property of the $\hat T({\bf R})$ is that they
commute with both $\hat H_L$ and $\Sigma_\omega({\bf r},{\bf r}^\prime)$ 
but not with $J_z$.  Since, moreover, within a Landau level there is no stable
subspace for all the $\hat T({\bf R})$, the eigenvalues 
$\sigma_\omega^{n,l}$ cannot depend on the angular momentum quantum number $l$.
At $\br \smeq 0$, Eq.~(\ref{eq:eigen_eq}), then, reads
\begin{equation}
 \sigma_\omega^{n,l} = \sigma_\omega^{n,0} =
  {1 \over \psi_{n,0}(0)} \int d{\bf r}\,\Sigma^{(0)}_\omega(|{\bf r}|)
     \psi_{n,0}({\bf r}) \; .
\end{equation}
Using the explicit expressions for $\Sigma^{(0)}_\omega({\bf r})$
[Eq.~(\ref{eq:sigma0})] and $\psi_{n,0}$,
we finally obtain for the eigenvalues
\begin{eqnarray} \label{eigen}
   \sigma_\omega^{n,l} &=& \sigma_\omega^{n,0} \\
     &=&{1\over 2}\int_{x_{\rm min}}^\infty dx\,{\exp\{-|m|x\}
     \over\sinh x}\exp\{-x^2/2\alpha^2\}L_n(x^2/\alpha^2)
     \nonumber
\end{eqnarray}
where $\alpha=L_H/L_T$ is the essential dimensionless parameter.
It is important to keep in mind that the screened
interaction already implicitly takes into account the effect of the
interaction on scales shorter than the Fermi wavelength, so the
integral over $x$ should be cut off for small $x$ at approximately
$x_{\rm  min}=1/(k_FL_T)$. 

\subsection{Reordering of the sum}

The interaction contribution to the thermodynamic potential is given
in terms of the eigenvalues $\sigma_\omega^n$ by
\begin{equation}
   \Omega={2 BA\over\phi_0}\,{1\over\beta}\sum_\omega\sum_{n=0}^\infty
         \ln(1+\lambda_0\sigma_\omega^n).
\end{equation} 
Here we have already taken proper account of the degeneracy of the
eigenvalues by the prefactor $2BA/\phi_0$. The magnetization per unit
area now follows by differentiation with respect to $B$,
\begin{equation}
  M=-{2 \over\phi_0 \beta}\sum_\omega\sum_{n=0}^\infty
     \left\{\ln(1+\lambda_0\sigma_\omega^n)+{B\lambda_0\over
     1+\lambda_0\sigma_\omega^n}{\partial\sigma_\omega^n\over\partial
     B}\right\}.
\end{equation}
When done naively, the sum over the quantum number in this
expression diverges. We assume that this is associated with the
inadequate treatment of the interaction at short distances. We expect
that when working with the full screened interaction, the contribution
of large quantum numbers is appropriately suppressed. Hence, we
reorder the sum in such a way that the sum becomes convergent and the
eigenvalues with sufficiently large $n$ do not contribute appreciably
to the sum.  This philosophy is completely analogous to the approach
taken in the work on the fluctuation contribution to the diamagnetic
susceptibility in superconductors above $T_c$ \cite{lee,eilenberger}. 
In fact, our reordering
closely follows the reordering proposed by Payne and Lee \cite{lee}.

In a first step, we compute
\end{multicols}
\lefthalfline
\begin{equation}
  B{\partial\sigma_\omega^n\over\partial B}=
     2\pi\int_0^\infty d\rho\,\rho\Sigma_\omega^{(0)}
   (\rho)\exp\left\{-{eB\over2\hbar}\rho^2\right\}
    {eB\over\hbar}\rho^2\left[L^\prime_n(eB\rho^2/\hbar)-{1\over2}
     L_n(eB\rho^2/\hbar)\right].
\end{equation}
To simplify this expression, we use the recursion relations for Laguerre
polynomials
\begin{eqnarray}
   xL_n^\prime(x)&=&nL_n(x)-nL_{n-1}(x)\\
   xL_n^\prime(x)&=&(n+1)L_{n+1}(x)-(n+1-x)L_n(x)
\end{eqnarray}
and obtain
\begin{equation}
    2B{\partial\sigma_\omega^n\over\partial B}=
     (n+1)[\sigma_\omega^{n+1}-\sigma_\omega^n]+n[\sigma_\omega^n-
     \sigma_\omega^{n-1}].
\end{equation} 
We can also rearrange
\begin{eqnarray}
  \sum_{n=0}^\infty(n+1)\ln{f_{n+1}\over f_n}
     &=&\sum_{n=0}^\infty n\ln f_n-\sum_{n=0}^\infty(n+1)\ln f_n
     \nonumber\\
     &=&-\sum_{n=0}^\infty\ln f_n.
\end{eqnarray}
Using these expressions, we have for the magnetization
\begin{eqnarray}
    M={1\over\phi_0}\,{1\over\beta}\sum_\omega\sum_{n=0}^\infty
     (n+1)\left\{2\ln{1+\lambda_0\sigma_\omega^{n+1}\over
     1+\lambda_0\sigma_\omega^n}-{\lambda_0[\sigma_\omega^{n+1}-
     \sigma_\omega^n]\over
     1+\lambda_0\sigma_\omega^n}-{\lambda_0[\sigma_\omega^{n+1}-
     \sigma_\omega^n]\over
     1+\lambda_0\sigma_\omega^{n+1}}\right\}.
\end{eqnarray}
In terms of the notation 
\begin{equation}
\label{xxx}
   X_\omega^n={\lambda_0[\sigma_\omega^{n+1}-\sigma_\omega^n]\over
     1+\lambda_0\sigma_\omega^n},
\end{equation}
we have
\begin{equation}
\label{m-final}
  M={1\over\phi_0}\,{1\over\beta}\sum_\omega\sum_{n=0}^\infty
     (n+1)\left\{2\ln(1+X_\omega^n)-X_\omega^n-{X_\omega^n\over
      1+X_\omega^n}\right\}.
\end{equation}
\righthalfline
\begin{multicols}{2}\noindent
For all cases considered below, $X^n_\omega\ll1$ so that
\begin{equation}
\label{magn-smallx}
   M \simeq -{1\over\phi_0}\,{1\over\beta}
   \sum_\omega\sum_{n=0}^\infty {(n+1)\over3}[X_\omega^n]^3.
\end{equation}
This expression will be our starting point for computing the
magnetization and the susceptibility. Referring back to the definition
of $X_\omega^n$ in Eq.\ (\ref{xxx}) above, we see that all
contributions to the susceptibility are at least third-order in
the interaction $\lambda_0$. 

\section{Magnetic susceptibility}
\label{sec:cal}

Using the general results of the last section, we now find expressions for the 
susceptibility in two limits-- small and large magnetic field-- and then 
evaluate the susceptibility in the intermediate regime numerically.
Before considering the various regimes,
it is useful to note that
$X^n_\omega$, as defined in Eq.~(\ref{xxx}), consists of two factors
with noticeably different behavior.  On the one hand, because the
integrand in Eq.~(\ref{eigen}) behaves as $1/x$ at small $x$, 
both $\sigma^n_\omega$ and $\lambda_0 / (1+\lambda_0\sigma_\omega^n)$
are dominated by a logarithmic singularity at zero and so have
little magnetic field dependence. On the other hand, using the relation
$L_n(x) - L_{n-1}(x) = - x L^1_{n-1}(x) / n$ ($L_n^1$ is a generalized
Laguerre polynomial), we can rewrite $\Delta \sigma^n_\omega \equiv
\sigma^{n+1}_\omega - \sigma^n_\omega$ as
\end{multicols}
\lefthalfline
\begin{equation} \label{eq:Delta_sigma}
  \Delta\sigma^n_\omega=-{1 \over (n+1)}{1\over2\alpha^2}
  \int_0^\infty dx\,x^2{e^{-|m|x}\over\sinh x}
   \exp\{-x^2/2\alpha^2\}L^1_n(x^2/\alpha^2).
\end{equation}
\righthalfline
\begin{multicols}{2}\noindent
Here the $x$ integration is well behaved at small $x$, and so 
the lower limit $x_{\rm min}$ can be replaced by zero.

\subsection{Small-magnetic-field (high-temperature) limit}

The small-magnetic field, or equivalently high-temperature, limit is defined
by $\alpha \!\gg\! 1$.  The factor $e^{-|m|x}/\sinh x$ provides an upper
cutoff at $\min(1,|m|^{-1})$ in the integrals Eqs.~(\ref{eigen}) and
(\ref{eq:Delta_sigma}). In addition, $x^2/\alpha^2$ is much smaller than one
in the entire range of integration.  We can therefore use the asymptotic
expression \cite{asymp_Lnalpa}
\begin{equation}
\label{bessel}
  e^{-x/2}L^\alpha_n(x)\simeq{\Gamma(\alpha+n+1)\over n!}(\nu x/4)^{-
  \alpha/2}J_\alpha([\nu x]^{1/2}),
\end{equation}
valid in the range $0 \!\leq\! x \!\leq\! n^{1/3}$ [$J_\alpha(x)$ denotes the
Bessel functions, $\Gamma(n)$ the Gamma function, and
$\nu=4n+2\alpha+2$] and obtain
\begin{equation}
\label{alln1}
  \sigma^n_\omega\simeq {1\over2}
   \int_{x_{\rm min}}^\infty dx{\exp(-|m|x)\over\sinh x}
     J_0\left(2\sqrt{n+1/2} \, x/\alpha\right)
\end{equation}
\begin{equation}
\label{alln2}
   \Delta\sigma^n_\omega\simeq-{1 \over \sqrt{n+1}}{1\over2\alpha}
    \int_0^\infty \!\! dx\,x{\exp(-|m|x)\over\sinh x}
     J_1\left( \frac{2\sqrt{n+1} \, x} {\alpha} \right) .
\end{equation}
For $n \!\ll\! n_0 \!=\! \alpha^2 \cdot \max(1,|m|)$,
Eqs.~(\ref{alln1}) and (\ref{alln2}) yield \cite{footnote1} 
\begin{equation}
\label{lown1}
   \sigma_\omega^n \simeq {1\over2}
   \int_{x_{\rm min}}^{\min(1,|m|^{-1})} {dx \over x} = {1\over 2}
         \ln(k_F L_T / {\rm max}\{1,|m|\})
\end{equation}
\begin{equation}
\label{lown2}
  \Delta\sigma^n_\omega \simeq -{1 \over 2\alpha^2}
  \int_0^\infty dx\,x^2{e^{-|m|x}\over\sinh x} \; ,
\end{equation}
up to constants of order one, and so the $n$ dependence can be neglected.
For $n \!>\! n_0$, both $\Delta\sigma^n_\omega$ and $\sigma^n_\omega$
depend on $n$.  However, for $\sigma^n_\omega$ the dependence
is only logarithmic, since it merely amounts to replacing the upper bound
of the integral by $\alpha/\sqrt{n}$. Hence the dominant $n$ dependence of
$X_\omega^n$ comes from $\Delta\sigma^n_\omega$.  

 From these results for the eigenvalues, the magnetization 
[Eq. (\ref{magn-smallx})] to lowest order in the small parameter 
$\ln^{-1}(k_FL_T)$ is
\end{multicols}
\lefthalfline
\begin{equation}
  M={1\over 3\phi_0}{1\over\beta}\sum_\omega{1\over\alpha^3\ln^3(k_FL_T)}
  \sum_{n=1}^\infty{1\over\sqrt{n}}\left\{ \int_0^\infty dx{x\exp(-|m|x)
  \over\sinh x}J_1\left(2\sqrt{n}x/\alpha\right)\right\}^3.
\end{equation}  
\righthalfline
\begin{multicols}{2}\noindent
The sum over $n$ converges only slowly and of order $n_0$ terms
contribute.  In view of the fact that $n$ is multiplied by
$x^2/\alpha^2$ in the argument of the Bessel function, we can replace
the sum over $n$ by an integral. This yields the final expression
\begin{equation} \label{eq:magn_smallB}
   M={C_T \over \pi \ln^3(k_FL_T)}(k_FL_T) \cl B
\end{equation}
where $\cl=e^2/12\pi m c^2$ and $C_T$ is given by
\begin{equation}
   C_T=\sum_{r=-\infty}^\infty\int_0^\infty{dn\over n^2}
        f_\omega^3(n) \simeq 0.97
\end{equation}
where we define
\begin{equation} \label{eq:function_f}
   f_\omega(n) \equiv \sqrt{n} \int_0^\infty dx{x\exp(-|m|x)
  \over\sinh x}J_1\left(2\sqrt{n}x\right) \; .
\end{equation}

We see that the interaction-induced contribution to the magnetization will
generally be larger than the Landau magnetization due to the large factor
$k_FL_T$ \cite{aslamazov}. The factor $1/\ln(k_FL_T)$ must be interpreted as
a renormalized interaction strength in the Cooper channel
\cite{aslamazov,doubleA}. It is interesting that the interaction
contribution to the susceptibility is {\it third order} in this renormalized
interaction strength. This unusual state of affairs can easily be understood
by considering the classical paths involved, as we shall discuss in
Section~\ref{sec:RG}.

The magnetic susceptibility $\chi$ obtained above is at zero field and to
lowest order in the renormalized interaction strength which is proportional
to $ 1/\ln(\kf L_T)$. It is possible to derive, with a similar approach, an
expression for $\chi(0)$ without expanding in the renomalized interaction 
strength.  This is done in Appendix~\ref{app}; one obtains
\begin{equation} \label{eq:chi0_ex}
  \frac{\chi(0)}{\cl} = \frac{3}{\pi} (\kf L_T) \, 
  \sum_\omega \int_0^\infty
   \frac{d\xi}{\xi^2} \lambda_\omega^3(\xi) f_\omega^3(\xi) 
\end{equation}
where we have introduced
\begin{eqnarray} 
        \lambda_\omega(\xi) & = &\frac{\lambda_0}{2+\lambda_0(2\sigma^0(T)  
       - g_\omega(\xi))}
        \label{eq:lambdaxi} \\
  g_\omega(x) & = & \int_0^\xi \frac{f_\omega(\xi')}{\xi'} d\xi'
        \label{eq:function_g} \\
        2\sigma^0 & \equiv &  2\sigma_\omega^n(B\!=\!0) = 
        \int_{x_{\rm min}}^\infty dx \frac{e^{-|m|x}}{\sinh x} 
        \nonumber \\
   & \simeq & \ln(\kf L_T / \max(1,|m|)) \; .
  \label{eq:sig0}
\end{eqnarray}

\subsection{Large-magnetic-field (low-temperature) limit}

In the high-magnetic-field, or equivalently the low-temperature, limit
defined by $\alpha\ll1$, the factor
$\exp\{-x^2/2\alpha^2\}L_n(x^2/\alpha^2)$ always cuts off the integral in
Eq.~(\ref{eigen}) at $x \!\ll\! 1$ so that we can approximate 
$\sinh x \simeq x$.  Hence, we find
\begin{eqnarray} \label{eq:hf_sigma}
   \sigma_\omega^n & \simeq & {1\over2}
   \int_{x_{\rm min}}^{\min [|m|^{-1},\alpha/\sqrt{n}]} {dx \over x} \\
   & = & {1\over2}\min[\ln(\kf L_H / \sqrt{n}),\ln(\kf L_T/|m|)] 
   \nonumber
\end{eqnarray}
\begin{equation}
  \Delta\sigma_\omega^n\simeq-{1 \over 2(n+1)}\int_0^\infty dy\,y
   e^{-\alpha|m|y-y^2/2}L^1_n(y^2).
\end{equation}
In this case, the sum over $n$ converges rapidly (faster than $1/n^2$), but
typically about $1/\alpha$ terms contribute to the Matsubara sum.
Neglecting again the logarithmic dependence of $\sigma_\omega^n$
on $n$ and $|m|$, we can make progress by noting that the sum over $m$ can
be turned into an integral. This yields the high-field result
\begin{equation} \label{eq:magn_largeB}
   M={ C_H \over\ \pi \ln^3(k_FL_H)}(k_FL_H)\cl B,
\end{equation}
where the constant $C_H$ is
\begin{equation}
   C_H=\int_{-\infty}^\infty dm \sum_{n=0}^\infty{(F_n(m))^3 \over (n+1)^2}
                        \simeq 0.74
\end{equation}
with
\begin{equation}
  F_n(m)=\int_0^\infty dy\,y e^{-|m|y-y^2/2}L^1_n(y^2).
\end{equation}
The principal difference between the results for high and low fields
is thus the replacement of the thermal length $L_T$ by
the magnetic length $L_H$. 

\subsection{Intermediate range}

When $\alpha$ is neither much smaller nor much larger than one, it is not
possible to obtain a simple expression for the magnetic response. In this
regime, we have performed a numerical integration of Eq.~(\ref{eigen}) to
compute the eigenvalues $\sigma^n_\omega$ as well as their derivatives with
respect to $B$. The magnetic susceptibility is then obtained through
the field derivative of Eq.~(\ref{m-final}),
\begin{equation} \label{eq:chi-final}
\chi = {-1 \over  \beta \phi_0} \sum_\omega \sum_n n
       {(X_\omega^{n-1})^2 dX_\omega^{n-1}/dB \over (1+X_\omega^{n-1})^2} \; ,
\end{equation}
by direct summation over the eigenvalue index and
the bosonic Matsubara frequencies.

Fig.~\ref{fig:harold1} shows the resulting $\chi/\cl$ as a function of
magnetic field $b=\alpha^{-2} = (2\pi L_T^2/\phi_0) B $ for fixed values of
the temperature. The cross-over between the low and high field regimes is
clearly seen.  As $B$ increases, $\chi$ has a slight maximum around
$L_H=L_B$ which arises from the competition between the increased field
sensitivity of large triangles and the thermal suppression of long sides.
In the large-field regime, the numerical result is in reasonable agreement
with the value obtained from the asymptotic expression
(\ref{eq:magn_largeB}):  for $L_T \smeq 10 L_H$ and $\kf L_H \smeq 64$,
$\chi/\cl \!\approx\!  0.28$ numerically and 0.18 analytically.

Fig.~\ref{fig:harold2} shows the temperature dependence of $\chi$ at fixed
$L_H/L_T$ in the low-field regime. The $(\kf L_T)/\ln^3(\kf L_T)$ behavior
is apparent, particularly in the inset. Again, the numerical result agrees
nicely with the asymptotic result:  at the low temperature $T/\Ef \smeq
10^{-3}$, Eq.~(\ref{eq:magn_smallB}) yields $\chi/\cl \!\approx\! 0.51 $
while our numerical result is 0.52.

\section{Semiclassical interpretation}
\label{sec:RG}

We have seen above that the magnetic response has a rather peculiar property:
it is {\it third-order} in the renormalized coupling constant $\tilde
\lambda = 2\ln^{-1}(\kf L_{T,H})$, both in the low and high field regimes.
Within the approach used up to now, it is difficult to understand the
physical origin of this behavior. In this section, we show that an
approach in terms of classical paths provides a natural understanding of
this fact. Considering for instance the low field regime (the argument
can be transposed to high fields with no essential difficulty), we
shall moreover recover precisely the expression Eq.~(\ref{eq:magn_smallB})
in a much simpler way.

 From the expression Eq.~(\ref{eq:sigma_sc}) for the particle-particle
propagator, it is clear that the interaction contribution to the
thermodynamic potential in Eq.~(\ref{omega_expanded}) can be written as a sum
over closed polygonal paths, where each vertex is associated with an
interaction event and the magnetic field enters only via the Aharonov-Bohm
factor associated with the magnetic flux enclosed by the polygon. Performing
the sum over bosonic Matsubara frequencies in Eq.~(\ref{omega_expanded}) and
grouping the field dependent terms together, one finds
\end{multicols}
\lefthalfline
\begin{equation}        
    \Omega = \sum_{n=1}^\infty \Omega^{(n)}  \label{omega_expanded2} \\
\end{equation}        
\begin{equation}        
     \Omega^{(n)} = -{1\over \beta} 
      \frac{(-\lambda_0)^n}{n} \, \int d{\bf r}_1 \ldots d {\bf r}_n \,
      \tilde \Sigma(\br_1, \br_2) \ldots \tilde \Sigma(\br_{n}, \br_1) 
      \coth\left( \frac{L_{\rm tot}(\br_1, \ldots, \br_{n}) }{ 2L_T} \right)
      \cos\left(\frac{4\pi A_{\rm tot}(\br_1, \ldots, \br_{n}) B }
                                     {\phi_0} \right) 
\end{equation}        
where $\tilde \Sigma(\br,\br')$ is the particle-particle propagator for 
$B \smeq 0$ and $\omega=0$, $L_{\rm tot}$ is the total length, and 
$A_{\rm tot}$ the area enclosed by the polygon.

As it stands, Eq.~(\ref{omega_expanded2}) is of little practical use because
the series in $n$ is strongly divergent (the term of order $n$ is typically
larger than the one of order $n-1$). It is, however, possible to apply a
simple renormalization-group argument, introduced in \cite{prl} and discussed
in more detail in \cite{long_paper}. Indeed, as we already stressed when
deriving the eigenvalues of the particle-particle propagator, $\tilde
\Sigma(\br,\br')$ must be cutoff at $\Lambda_0 = 1/\kf$ because the use of
the screened interaction assumes that all high-momentum degrees of freedom
have already been integrated out \cite{doubleA}. We therefore have
\begin{eqnarray}
        \tilde \Sigma(\br,\br';\Lambda_0) & = & 
                \frac{1}{4\pi L_T |\br-\br'| \sinh(|\br-\br'|/L_T)}  
                \qquad \mbox{for $|\br-\br'| > \Lambda_0$} \label{eq:hsigma}\\
                & = & 0 \qquad\qquad \qquad\qquad \qquad
   \qquad\qquad \mbox{for $|\br-\br'| < \Lambda_0$} \; .
          \nonumber
\end{eqnarray}
Refs. \cite{prl,long_paper} show that if a new length scale $\Lambda >
\Lambda_0$ much smaller that any other characteristic length scale of the
problem ($L_T$ or $L_H$) is introduced, one can replace $\Lambda_0$ by
$\Lambda$ in Eq.~(\ref{eq:hsigma}) provided the ``bare'' coupling constant
$\lambda_0$ in Eq.~(\ref{omega_expanded2}) is replaced by the renormalized
one
\begin{equation}
        \lambda_{RG}(\Lambda) = \frac{\lambda_0}
          {1 + (\lambda_0/2) \ln (\Lambda/\Lambda_0)}.
\end{equation}

For the leading behavior in $\ln^{-1}(\kf L_T)$, we can assume $\Lambda =
\epsilon L_T$ with $\epsilon \ll 1$ but assumed fixed as $\ln(\kf L_T)$ goes
to infinity. In that case $\lambda_{RG}(\Lambda) = 2\ln^{-1}(\kf L_T) + O
(\ln^{-2}(\kf L_T))$ is small, and Eq.~(\ref{omega_expanded2}) becomes a
genuine perturbative expansion whose leading behavior is given by the first
non-vanishing term. Clearly, $\Omega^{(1)}=0$, and $\Omega^{(2)}$ is
independent of the magnetic field and so does not contribute to the magnetic
response. Therefore, the leading behavior is given by $\Omega^{(3)}$-- that
is, {\it third-order} in $\ln^{-1}(\kf L_T)$-- as illustrated in
Fig.~\ref{fig:traj}, and we have
\begin{equation}
\chi(B\!=\!0)  =  
{1 \over 3 \beta} \left(\frac{2}{\ln(\kf L_T)} \right)^3 
   \int_{L_2, L_3, L_{23} > \Lambda} d\br_2  d \br_3 \,
   \tilde\Sigma({\bf 0},\br_2) \tilde\Sigma(\br_2,\br_3) 
   \tilde\Sigma(\br_2,{\bf 0})
                 \coth \left( \frac{L_{\rm tot} }{ 2L_T} \right)
                 \left( \frac{4\pi A_{\rm tot}}{ \phi_0} \right)^2 
\end{equation}
\righthalfline
\begin{multicols}{2}
($L_{2}=|\br_2|$, $L_{3}=|\br_3|$,  $L_{23}=|\br_3-\br_2|$).  
Expressing all distances in the integral in units of $L_T$ gives
\begin{equation} \label{eq:chi_RG}
 \chi(B\!=\!0) = \frac{C_{RG}}{\pi \ln^3(\kf L_T)} (\kf L_T) \cl
\end{equation}
with the constant $C_{RG}$ given by 
\begin{equation}
        C_{RG} = \frac{1}{\pi^2}  \int d\br_2  d \br_3 \,       
  \frac{1}{L_{2}L_{23}L_{3}} 
        \frac{A^2_{\rm tot} \coth(L_{\rm tot} / 2)}
        {\sinh L_{2} \sinh L_{23} \sinh L_{3}} \; .
\end{equation}
Because of the factor $A_{\rm tot}^2$ in the numerator, the integrand here is
regular and the cutoff can be taken to zero. By considering the limit
$\lambda_0 \ll \ln(\kf L_T) \ll 1$ (instead of $\lambda_0=1$) where both the
standard and renormalization group approaches are accurate, it can be shown
that $C_{RG}=C_T$, so that Eq.~(\ref{eq:chi_RG}) is strictly equivalent to
Eq.~(\ref{eq:magn_smallB}). {\it This approach shows clearly that the third
power of the coupling constant arises because only trajectories with three or
more vertices enclose flux.}

\section{Discussion and summary}
\label{sec:con}

In this paper, we study the interaction contribution to the magnetization of
a two-dimensional electron gas in the limit $L_T\ll R_c,\ell_{\rm el}$. We
find that this interaction-induced contribution is paramagnetic for the
repulsive Coulomb interaction and dominates over the Landau diamagnetism at
small enough temperatures and fields \cite{aslamazov}.  The Pauli
paramagnetism is even smaller than Landau diamagnetism in GaAs/AlGaAs
heterostructures because of the small effective mass and the reduction in
the $g$ factor. It appears from the quantitative answer that one needs to go
to rather low temperatures before the interaction becomes truly larger than
the Landau susceptibility (cf.\ Fig.\ \ref{fig:harold2}).  Still, such
temperatures are possible in two-dimensional electron gas systems.  It
should be possible to distinguish the interaction contribution by way of
either its temperature dependence (since the Landau susceptibility is $T$
independent) or its dependence on $\kf$ in a gated structure.

We find that the leading contribution to the magnetization is third-order in
the renormalized Coulomb interaction. This can be given a natural
semiclassical interpretation in terms of the classical-path picture for the
thermodynamic potential: a polygon must have at least three sides in order to
enclose area. Moreover, the much simpler classical-path approach gives
exactly the same answer as the eigenvalue calculation. Higher-than-third
order contributions (in the bare coupling constant) predominantly lead to an
substantial downward renormalization of the coupling constant. This picture
has been made precise in the present paper by means of a
renormalization-group approach.

In the low-field (high-temperature) limit $L_T \!\ll\! L_H$, the temperature
dependence of the susceptibility is $1/T\ln^3T$: this comes from the thermal
length $L_T$ which dominates here because only trajectories shorter than
$L_T$ make significant contributions to the Green functions. At the lowest
temperatures, this behavior is cut off by a finite magnetic field once $L_H
\!<\! L_T$. In the high-field (low-temperature) limit $L_T\gg L_H$, the
susceptibility is no longer temperature dependent. The Green function is
still dominated by trajectories shorter than $L_T$, but now trajectories
enclosing more area than $L_H^2$ contribute with random signs due to the
Aharonov-Bohm phases. Hence, in this case, the relevant cutoff length is
$L_H$.

So far, we ignored dephasing due to inelastic scattering. At low
temperatures, this should be mostly due to electron-electron
scattering. We expect however that dephasing will not significantly
affect our results. Within the semiclassical approach
employed in this paper, dephasing suppresses the contribution of
trajectories longer than the dephasing length $L_\phi$. For a clean
Fermi liquid such as discussed here, one expects $L_\phi\sim 1/T^2$.
Thus, at sufficiently low temperatures the dephasing length should
always be longer than the thermal length $L_T\sim 1/T$.
Correspondingly, the suppression of trajectories due to thermal
smearing should always set in before the suppression due to dephasing.

It is interesting to compare the present results with the contributions to
the susceptibility of (chaotic) mesoscopic samples of linear size $L$ within
the independent-electron approximation \cite{shapiro,oppen,ullmo}
\begin{equation}
  \chi\sim\cl,
\end{equation}
and due to interactions \cite{prl}
\begin{equation}
  \chi\sim{(k_FL)\over \ln(k_FL)}\cl,
\end{equation}
where we have taken $L_T \!\sim\! L$, $T \!>\! \Delta$ ($\Delta$ is the level
spacing), and $L_H \!\gg\! L$. In contrast to the bulk results derived in
this paper, these expressions are zeroth or first order in the renormalized
interaction constant. These contributions exist for mesoscopic samples
because, in finite-size systems, flux-enclosing trajectories are produced by
scattering from the geometric boundaries of the system. Nevertheless, apart
from the different order in the renormalized interaction, the finite-size
result due to interactions is qualitatively the same as the bulk result
derived here.

Orbital magnetism in mesoscopic samples has been a controversial issue over
the last decade, both for ballistic and diffusive structures, ring and dot
geometries \cite{Levy2,Benoit,Levy1,Webb1,Webb2,Efetov}. 
In particular, the fact that the measured values are apparently
substantially larger than the theoretical results has attracted a lot of
attention. In order to benchmark the theory in a simpler system, we think
the magnetization of a clean two-dimensional electron gas should be measured
and that this would provide valuable information in addressing the
``persistent current problem'' in rings and dots.

\acknowledgments 
We thank Rodolfo Jalabert and Klaus Richter for valuable discussions.  The
LPTMS is ``Unit\'e de recherche de l'Universit\'e Paris~11 associ\'ee au
C.N.R.S.''. FvO was partly supported by SFB 341 (K\"oln-Aachen-J\"ulich).

\appendix

\section{}

\label{app}

In this  appendix, we derive  an expression for the  susceptibility at
zero field, Eq.~(\ref{eq:chi0_ex}), which  does not involve expanding in
the renormalized  coupling constant.  Starting from  
the magnetization  Eq.~(\ref{magn-smallx}), the essential ingredient
needed is expressions for the eigenvalues. One can check that both the
Bessel   approximations   Eqs.~(\ref{alln1})-(\ref{alln2})   and   the
replacement  of the discrete  sum over  Landau level  index $n$  by an
integral  only  yield  corrections  of  order $B^2$  to  the  magnetic
susceptibility. Therefore,  as long as  we are only interested  in the
susceptibility at $B \smeq 0$, we  can make the change of variables $n
\to \xi  = b  n$ where $b  = \alpha^{-2}  = (2\pi L_T^2/\phi_0)  B$ is
proportional to the magnetic field.  We can thus write
\begin{equation}
       2 \Delta \sigma^{n-1}_\omega  =  - b f_\omega(\xi) / \xi 
\end{equation}
where $f_\omega(\xi)$ is defined by Eq.~(\ref{eq:function_f}).  In the same 
way, taking the derivative of Eq.~(\ref{alln1}) with respect to 
$b$ yields
\begin{equation}
       2 \frac {d \sigma^{n-1}_\omega}{db}  =  - \frac{1}{b} f((n-1/2)b)
        \simeq - \frac{1}{b} [f(\xi)  + O(b)].
\end{equation}
Thus, using the notation of Eqs.~(\ref{eq:function_g}) and (\ref{eq:sig0}),
\begin{equation}
        2\sigma^{n-1}_\omega  =  2\sigma^0        - g((n-1/2)b) 
   \simeq 2\sigma^0 -  g(\xi) + O(b) \; .
\end{equation}

 From these expressions for the eigenvalues, we obtain 
\begin{eqnarray}
        X^{n-1}_\omega = X_\omega(\xi,b)  & = &  
                          -b \frac{ \lambda_0 f_\omega(\xi) / \xi }
                    {2 + \lambda_0 (2\sigma^0 - g(\xi))} + O(b^2) 
              \nonumber \\
        & = & -b \lambda_\omega(\xi) f_\omega(\xi) / \xi   +  O(b^2) .
\end{eqnarray}
Note  that   it  is  necessary  to  compensate   the  factor  $b^{-2}$
originating from the  term $n dn$ when changing  variables from $n$ to
$\xi$.  Using  the  above  expression,  one can  therefore  write  the
magnetization up to corrections of order $b^2$,
\begin{equation}
   M = - {1\over 3 \phi_0 \beta}
  \sum_\omega   \int   \frac{\xi   d\xi}{b^2}   
        X_\omega(\xi,b)^3 + O(b^2)  \;  ,
\end{equation} 
which immediately gives Eq.~(\ref{eq:chi0_ex}).

\end{multicols}

\eject

\begin{figure}
\begin{center}
\leavevmode
\epsfxsize=8.5cm
\epsfbox{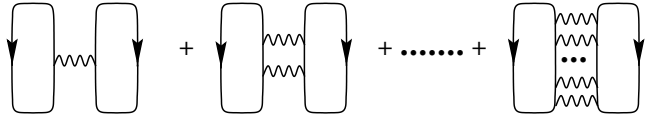}
\end{center}
\caption{
Leading Cooper-channel diagrams for the interaction contribution to the
thermodynamic potential. Because we can take the interaction to be local
(a $\delta$-function), the corresponding Fock-like diagrams differ from the
Hartree-like diagrams shown only by a factor of $-1/2$.
}
\label{fig:cooper}
\end{figure}

\begin{figure}[tb]
\begin{center}
\leavevmode
\epsfxsize = 12.0cm
\epsfbox{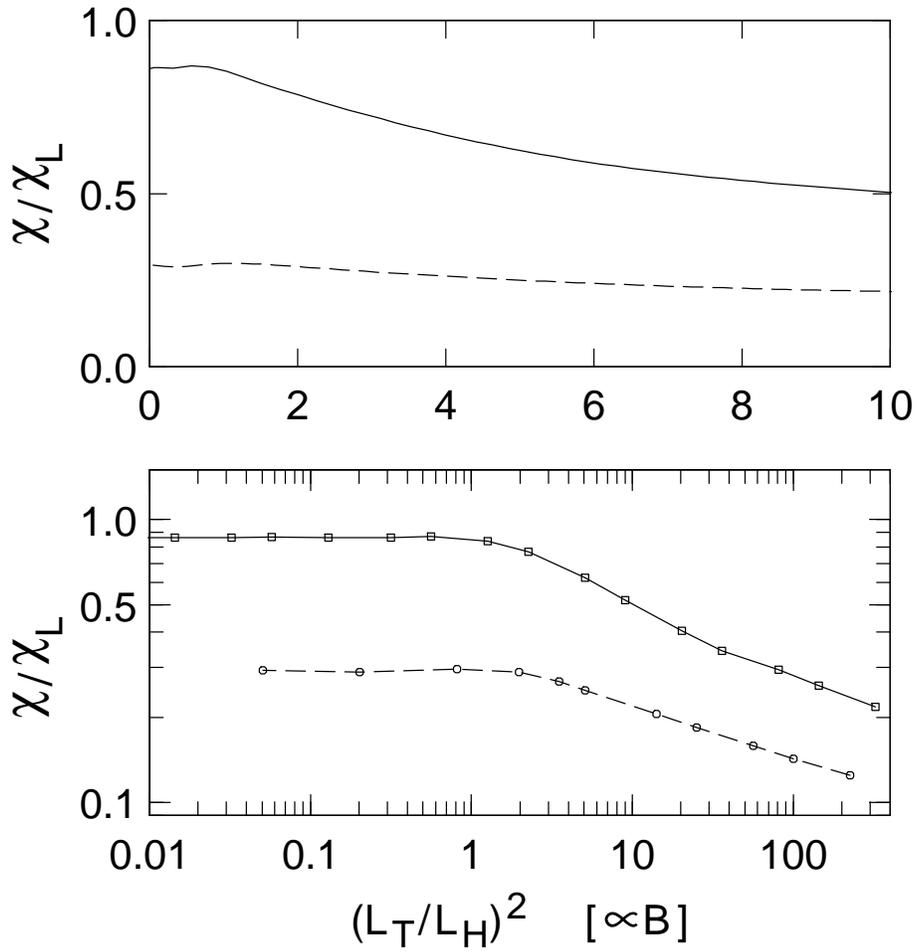}
\end{center}
\caption{
Magnetic field dependence of the susceptibility at two temperatures, $E_F/T
\smeq 2000$ (solid) and $200$ (dashed). The Landau susceptibility $\chi_L$ is
the natural unit for $\chi$; the lower panel shows the same data on an
logarythmic scale. Note the cross-over in behavior when $L_T \!\approx\! L_H$.
}
\label{fig:harold1}
\end{figure}

\begin{figure}[t]
\begin{center}
\leavevmode
\epsfxsize = 12.0cm
\epsfbox{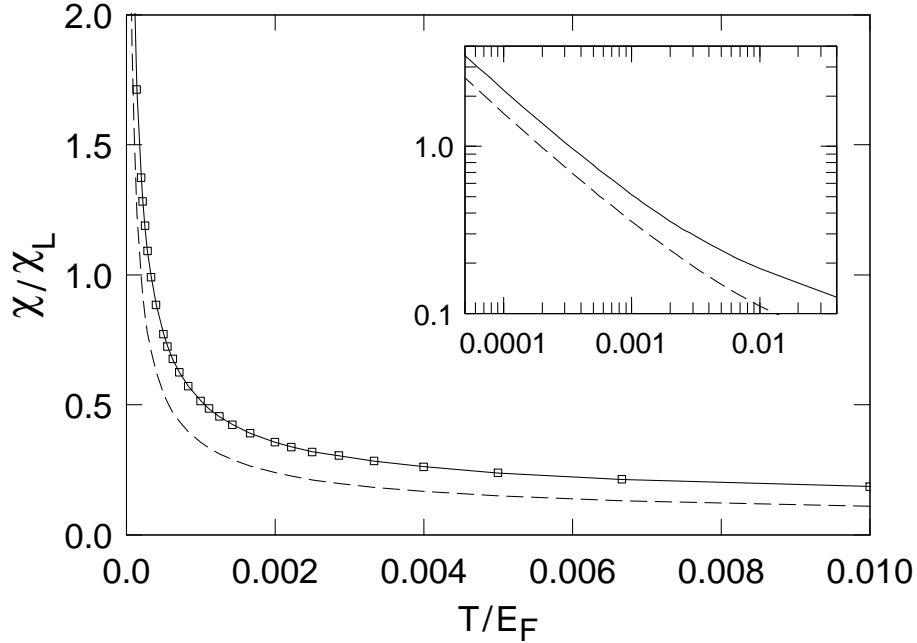}
\end{center}
\caption{
Temperature dependence of the interaction-induced low-field susceptibility;
$L_H/L_T \smeq 8.9$ is fixed. The total susceptibility differs from the 
interaction-induced contribution shown here only by a {\it constant offset} 
due to the Landau and Pauli contributions. The inset shows the same data on 
a log-log scale. For this low field, the contribution of the $\omega \smeq 
0$ Matsubara frequency (dashed) gives a substantial portion of the result 
(solid). Note the approximate power law increase in $\chi$ at low 
temperatures, consistent with the asymptotic expression $1/T\ln^3T$.
} 
\label{fig:harold2} 
\end{figure}

\begin{figure}
\begin{center}
\leavevmode
\epsfxsize=6.0cm
\epsfbox{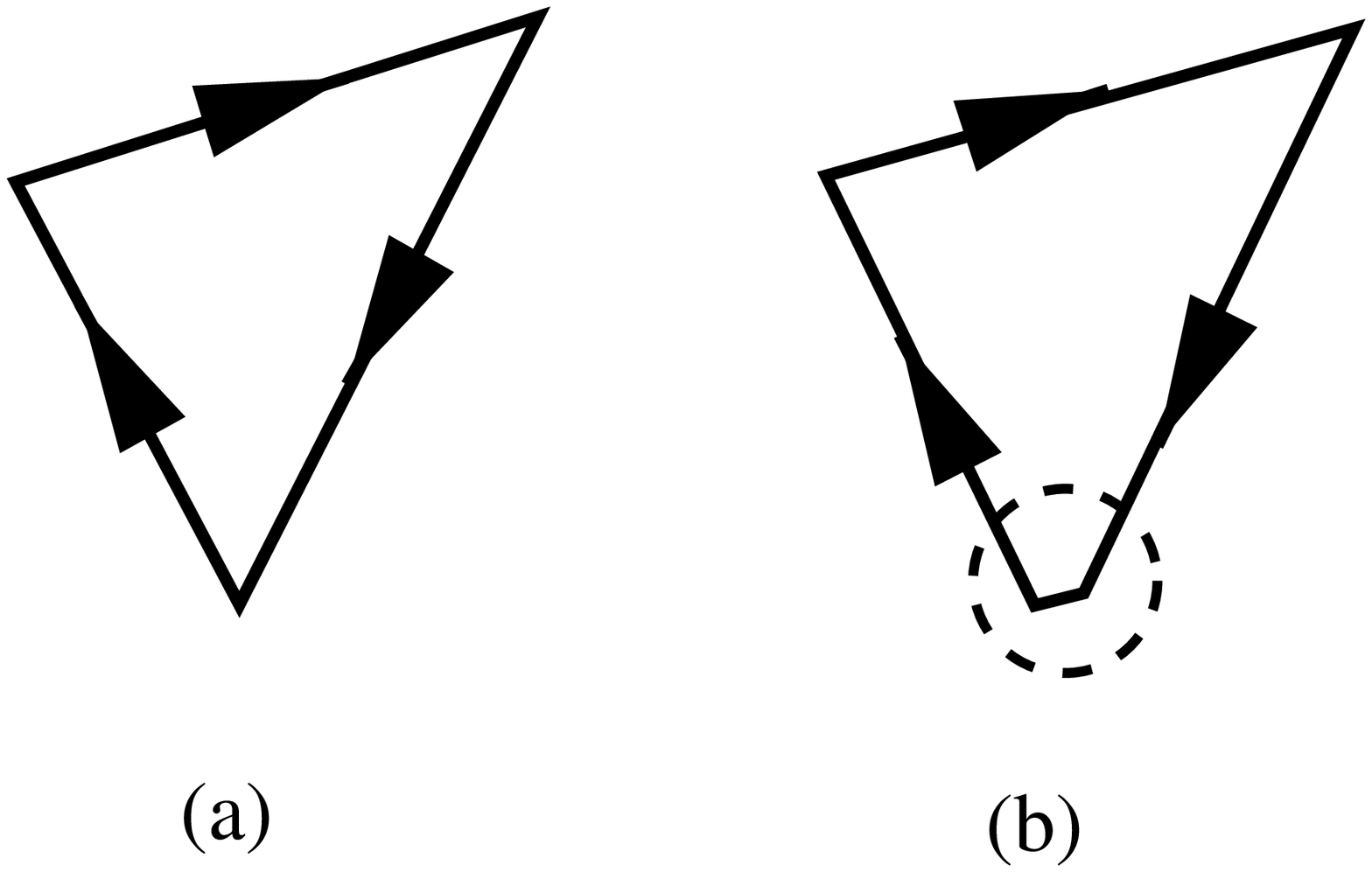}
\end{center}
\caption{
(a) Typical trajectory of lowest order in the coupling constant which
contributes to the interaction contribution to the magnetization. 
Each vertex of the polygon corresponds to an interaction event. 
Note that at least three interaction events are required to 
obtain a trajectory enclosing magnetic flux. This explains the fact that
the interaction contribution to the magnetization is third-order 
in the (renormalized) interaction. (b) Higher-order contributions 
predominantly lead to a renormalization of the third-order result 
due to short trajectories like the one shown. 
}
\label{fig:traj}
\end{figure}


\begin{references}
\bibitem{footnote1} Strictly speaking, the Bessel function
  approximation Eq.\ (\ref{bessel}) and hence Eqs.\ (\ref{alln1}) and
  (\ref{alln2}) are valid only for large $n$. However, one 
  checks by direct computation that in the present case, no serious
  errors are incurred when using Eq.\ (\ref{bessel}) even for small
  $n$.
\end{references}
\end{document}